\documentclass[journal]{IEEEtran}
\usepackage{cite}

\usepackage{graphicx}
\usepackage{amsmath}
\usepackage{multirow}
\usepackage{optidef}
\usepackage{bbm}
\usepackage{mathabx}
\usepackage{bm}
\usepackage{xcolor}
\usepackage{pdfpages}

\ifCLASSINFOpdf
\else
\fi
\usepackage{array}
\usepackage{fixltx2e}
\begin{document}
%
\title{Distribution LMP-based Transactive Day-ahead Market with Variable Renewable Generation}
%
%
%

\author{M. Nazif Faqiry,
        Lawryn Edmonds,
         Hongyu Wu
\thanks{M. Nazif Faqiry, Lawryn Edmonds, and Hongyu Wu are with the department of Electrical and Computer Engineering, Kansas State University.} 
}

%
%

\markboth{}%
{Shell \MakeLowercase{\textit{et al.}}: Bare Demo of IEEEtran.cls for IEEE Journals}
%



\maketitle

\begin{abstract}
The large-scale penetration of variable renewable energy (VRE) and their generation uncertainties poses a major challenge for the distribution system operator (DSO) to efficiently determine the day-ahead real and reactive power distribution locational marginal prices (DLMPs) and their underlying components. In this paper, we propose a DLMP-based transactive day-ahead market (DAM) model, that in addition to energy and losses, determines prices for creating congestions and voltage violations under peak-load and large-scale stochastic VRE penetration conditions. To account for the VRE uncertainties and the effect of their large-scale penetration on the DLMP components and distributed energy resources' (DERs) schedules, we propose a novel data-driven probability efficient point (PEP) method that computes the optimal total VRE generation at different confidence (risk) levels to incorporate in the proposed transactive DAM model. We perform a wide range of simulation studies on a modified IEEE 69-node system to validate the proposed methods and demonstrate the effect of peak load conditions, large-scale VRE penetration, and inclusion of battery energy storage systems (BESSs) on the resulting positive or negative real and reactive power DLMPs and their components.
\end{abstract}
\IEEEpeerreviewmaketitle
\section{Nomenclature}
\begin{tabbing}
  XXXXXXXX \= \kill
  $i,t,w,s$ \>Index of node, time, segment, and sample \\
  $T, N$ \> 	Number of timeslots, nodes/lines \\
  $G,D,B,R$ \> Generation, demand, battery, renewable scripts \\
  $E,L,V,C$ \> Energy, loss, voltage, congestion superscripts \\
  $p,q,V,\delta$ \> Real, reactive power injection, voltage, angle \\
  $P,Q,L^P,L^Q$ \> Real, reactive power flow, real, reactive losses \\
  $u(\cdot),d\{\cdot\}$ \> Immediate upstream node, downstream subtree \\
  $\mathrm{r},\mathrm{x}$ \> Resistance, reactance \\
  $\lambda,\mu,\rho$ \> Shadow price of energy, voltage, congestion \\
  $\pi,\gamma,(\cdot)^{\gamma}$ \> Probability, confidence level, $\gamma$-efficient point \\
  $\mathbf{u},\mathbf{v}$ \> Realizations of a stochastic vector \\
  $\bm{\mathcal{U}},\bm{\mathcal{D}}$ \> Upstream, downstream matrices \\
  $\mathcal{O},\Omega$ \> Objective function, DLMP  \\
  $\kappa$ \> Reactive to real power ratio $ q/ p$ \\
  $\mathrm{c}, \zeta, y $ \> Bid, value of lost load/generation, curtailment \\
  $\mathbbm{z}, \mathbbm{M}$ \> Binary variable, set of integer constraints\\
  $e, \mathcal{E}$ \> Battery energy, state of charge \\
  $\check{\mathbf{\cdot}},\hat{\mathbf{\cdot}}$ \> Indicates injection, extraction \\
  $\tilde{\mathbf{\cdot}},\underline{\mathbf{\cdot}},\overline{\mathbf{\cdot}}$ \> Indicates stochastic, lower, upper limit\\
 \end{tabbing}
\section{Introduction}
%
%
%
%

\IEEEPARstart{W}{ith} the ever-increasing penetration of variable renewable energy (VRE) generation and the emerging market participation of other distributed energy resources (DERs) such as price responsive loads (RLs), conventional generators (CGs), and battery energy storage systems (BESSs), pricing based on distribution locational marginal price (DLMP) has acquired considerable research attention as the means to establish the true price of electricity in a transactive energy market \cite{1,2,3,4,5}. Distribution system operators (DSOs) are expected to leverage the availability of services from DERs in a day-ahead market (DAM) and capture value by optimizing for least cost operation while respecting the distribution grid's constraints \cite{6}. One of the key challenges for efficient energy transaction mechanisms is designing them in a way to motivate active participation of customers owning DERs \cite{7}. A traditional way to achieve this is based on locational marginal pricing (LMP) currently practiced in the wholesale market.
	
Unfortunately, due to the lower reactance/resistance ($\mathrm{x/r}$) ratio in the distribution system, the LMP-based transmission system DAM is not a suitable model in the distribution system as it incorporates DC optimal power flow (DCOPF)--resulting in LMPs that do not reflect system losses, voltage violations, and distribution lines' congestion. A more detailed pricing model that reveals these additional components is required to obtain real and reactive DLMPs and their components under low and high level DER penetration that may yield reverse power flow in a radial distribution grid. The existing literature on DLMP pricing mostly focuses on production cost modeling and excludes participation of the RLs and zero-marginal-cost VREs while ignoring the uncertainty involved in their generation. To the best of our knowledge, a detailed study that accounts for the impact of large-scale VRE penetration on the resulting component-wise positive or negative real and reactive power DLMPs does not exist in the literature.

In this paper, we aim to address the aforementioned inadequacies. We develop a mixed integer linear programming (MILP) model for the DSO's transactive DAM considering different types of DERs, including stochastic VRE generation, and compute DLMP values and its components. We incorporate a detailed linearized power flow (LPF) to set up linear constraints for the DAM social welfare maximization problem. We decompose the DLMP into its energy, loss, voltage violation, and congestion components by providing comprehensive sensitivity analyses. To account for resource availability due to uncertainty in VRE generation, we propose the novel data-driven probability efficient point (PEP) method to compute VRE generation amounts based on historical data at different confidence or risk levels. Moreover, we provide interesting results on the significance of reactive power pricing and present component-wise details of positive and negative DLMP components under different confidence levels in the PEP method and high-scale VRE penetration. 

\subsection{Related Work}
Several papers have attempted to introduce DLMP-based pricing or its market models for the distribution system either through the use of DCOPF or a variant of AC optimal power flow (ACOPF). DCOPF-based DLMP formulations are presented in \cite{1,8,9,10}. Because of the lower $\mathrm{x/r}$ ratio in the distribution system, the use of DCOPF has been shown to introduce significant errors \cite{2}. Moreover, these formulations lack inherent features to include losses, voltage violations, and reactive power pricing that are required in a transactive distribution market. References \cite{1} and \cite{9} propose DLMP-based methods through quadratic programming and chance-constrained mixed integer programming that use DCOPF to define line congestions and alleviate it through dynamic tariffs. In \cite{8}, the authors propose a transactive DAM model in which the DSO interacts with the ISO and DERs in a local distribution area to participate in demand response programs--focusing more on the interaction of the market players while assigning equal DLMPs to DERs. Reference \cite{10} proposes two benchmark pricing methodologies, namely DLMP and  iterative DLMP, for a congestion-free energy management of the distribution grid through aggregators that have contractual terms with buildings to decide their reserve and energy schedules by interacting and providing flexible demand to the DSO. 

Recent studies that are related to our work, and model a variant or approximation of ACOPF in the DLMP formulation and determine one or more of its components, are reported in \cite{2,3,4}. In \cite{2}, a novel LPF method has been introduced that derives energy and loss components of the DLMP while disregarding voltage violation and congestion components, participation of RLs, and the uncertainty in the VRE generation. In \cite{3}, a DAM model based on DLMP has been proposed to manage congestion and provide voltage support by using mixed-integer second-order-cone programming to model ACOPF and determine binary variables such as feeder configuration status and tap locations of shunt capacitors and transformers. While only optimizing for least cost generation, this study does not consider VRE, RLs, or BESS units with temporal constraints. Similarly, in \cite{4}, authors use a trust-region based solution methodology to obtain DLMP and its constituents through a first-order approximation of the AC power flow manifold model in \cite{11} showing better performance over \cite{2}, however, it does not study inclusion of BESSs and large-scale penetration of stochastic VRE that may result in reverse power flow and negative real and reactive DLMPs. 

\subsection{Technical Contributions}
\noindent ($\textbf{\textit{i}}$)	We propose a complete DLMP-based transactive DAM model for the DSO to schedule generation resources, including VRE, reliably and economically. We derive and incorporate two additional DLMP components, namely voltage violation and congestion prices, and build upon the recent work in \cite{2}.   \\ 
($\textbf{\textit{ii}}$)	In the proposed transactive DAM, to account for the uncertainty in the VRE generation, we implement a novel data-driven MILP based on a large amount of historical VRE data and then optimally solve PEPs without assuming any probability distribution function.\\ 
($\textbf{\textit{iii}}$) We demonstrate, for the first time, the component-wise significance of real and reactive power pricing in the proposed DLMP-based transactive model and analyze the impact of increasing VRE on individual DLMP components. We also highlight novel observations on negative real and reactive DLMP components under high VRE penetration.\\ 
\indent We organize the rest of this paper as follows. In Section III, we present the system model in three subsections. We describe the transactive DAM problem formulation and the DLMP decomposition in Section IV, present simulation results in Section V, and conclude in Section VI.

\section{System Model}
We assume a radial, symmetric, and balanced distribution network. Similar to our previous work in \cite{12}, we index the root node as 0 and order the rest of the system nodes from $1$ to $N$. This allows us to devise the same index to lines and nodes from $1$ to $N$ by excluding the root (substation) node.  It turns out that this way of indexing in a radial distribution system greatly simplifies the presentation of the power flow model and the sensitivity analyses. In Fig. 1, $i$ is the node that is connected to its immediate upstream node $u(i)$ and a set of down-stream nodes $d\{i\}$. Line $i$ is the line that connects node $u(i)$ to $i$. All quantities related to line $i$ are subscripted by index $i$. For instance, line resistance, reactance, voltage drop, and angle difference caused in voltage are denoted by $\mathrm{r}_i$, $\mathrm{x}_i$, $\Delta V_i$, and $\Delta \delta_i$ and real and reactive power flows and losses are shown by $P_i$, $Q_i$, $L_i^P$, and $L_i^Q$. Similarly, the real and reactive power injections at node $i$ are denoted by $p_i$ and $q_i$.
\begin{figure}[!htb]
\begin{center}
\includegraphics[width=3.5in]{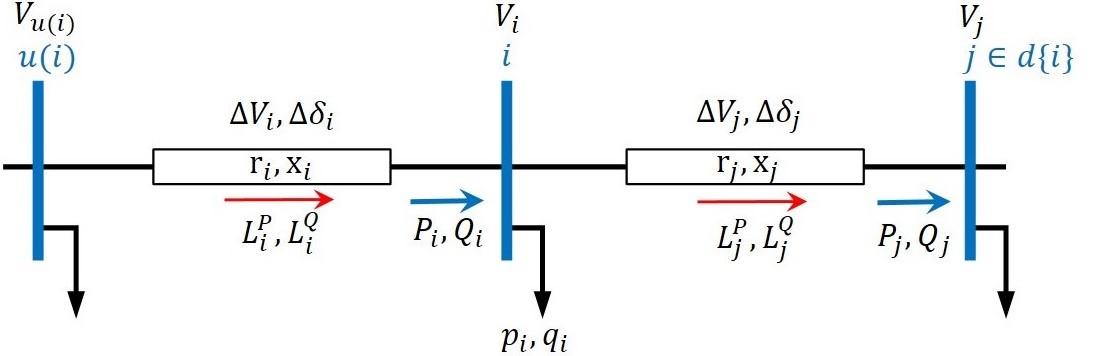}
\caption{Radial branch with line $i$ connecting bus $u(i)$ to $i$.}
\label{fig1}
\end{center}
\end{figure}
\setlength{\textfloatsep}{6pt plus 1.0pt minus 2.0pt}
\subsection{Power Flow Model}
The real and reactive power injection at each node is modeled by (1) with stochastic $\tilde{p}_{i,t}^R$ and $\tilde{q}_{i,t}^R$. Although load is also stochastic in nature, we only consider the VRE generation's stochasticity. However, the PEP method introduced later can be easily extended to account for the demand uncertainty.
\begin{subequations}
\begin{align}
p_{i,t} &=  p_{i,t}^D - p_{i,t}^G - p_{i,t}^B - \tilde{p}_{i,t}^R \label{eq1a} \\
q_{i,t} &=  q_{i,t}^D - q_{i,t}^G - q_{i,t}^B - \tilde{q}_{i,t}^R. \label{eq1b} 
\end{align}
\end{subequations}
In the RHS of (1), $p_{i,t}^D$,$q_{i,t}^D$ and $p_{i,t}^G$,$q_{i,t}^G$ relate to constraints given by the sets $\mathcal{D}(x_t )$, $\mathcal{G}(x_t )$ \cite{13}. The third term, $p_{i,t}^B,q_{i,t}^B$ relates to BESS constraints in Section III.C. The fourth term, $\tilde{p}_{i,t}^R,\tilde{q}_{i,t}^R$, is determined by the PEP method in section III.B.
\begin{flalign}
&\underline{\mathcal{G}(x_t )}                                                                                   &\mid     \qquad              &\underline{\mathcal{D}(x_t )} \nonumber\\
& p_{i,t}^G=\sum_{w}p_{i,w,t}^G                                                                            &\mid      \qquad               & p_{i,t}^D=\sum_{w}p_{i,w,t}^D \label{eq2}\\
&0\leq p_{i,w,t}^G\leq \overline{P}_{i,w}^G                                                             &\mid      \qquad              & 0\leq p_{i,w,t}^D\leq \overline{P}_{i,w}^D \label{eq3}\\
&\underline{P}_{i}^G\leq p_{i,t}^G\leq \overline{P}_{i}^G                                       &\mid     \qquad                & \underline{P}_{i}^D\leq p_{i,t}^D\leq \overline{P}_{i}^D \label{eq4}\\
&-\kappa_{i}^G\cdot p_{i,t}^G \leq q_{i,t}^G\leq \kappa_{i}^G\cdot p_{i,t}^G     &\mid      \qquad               & q_{i,t}^D=\kappa_{i}^D\cdot p_{i,t}^D. \label{eq5}
\end{flalign}
In (\ref{eq2}), the sum of segment generation and load should equal the total load and generation. The segment generation and load has to lie within the given bounds (\ref{eq3}) and the total real and reactive power generation and load have to satisfy (\ref{eq4}) and (\ref{eq5}). 

To establish the relationship between real and reactive power injections in (1) with the node voltages and line losses, we adopt the LPF method in \cite{2} and modify it to serve our model. We first derive the relationship between voltage and power injections and then consider losses and congestion. Using the indexing described earlier, we rewrite LPF, given by (6)-(7), based on voltage drops and its angle differences. According to Fig. 1, the power flow equations can be written as,
\begin{equation}
\resizebox{1\hsize}{!}{$(P_i + L_i^P) + j(Q_i + L_i^Q)=\frac{V_{u(i)}^2 - V_{u(i)}V_i \mathrm{cos}(\Delta\delta_i)) - j V_{u(i)}V_i \mathrm{sin}(\Delta\delta_i))}{\mathrm{r}_i + \mathrm{x}_i}. \label{eq6}$} 
\end{equation}
Expanding (\ref{eq6}) to real and imaginary components yield,
\resizebox{1\hsize}{!}{$(P_i + L_i^P) =\frac{\mathrm{r}_i \cdot \big[ V_{u(i)} \big( V_{u(i)}  -  V_i \mathrm{cos}(\Delta\delta_i)  \big)  \big]}{\mathrm{r}_i^2 + \mathrm{x}_i^2}   +  \frac{\mathrm{x}_i \cdot \big( V_{u(i)} V_i\mathrm{sin}(\Delta\delta_i)  \big)}{\mathrm{r}_i^2 + \mathrm{x}_i^2}$} 
\resizebox{1\hsize}{!}{$(Q_i + L_i^Q) =-\frac{\mathrm{r}_i \cdot \big( V_{u(i)} V_i\mathrm{sin}(\Delta\delta_i)  \big)}{\mathrm{r}_i^2 + \mathrm{x}_i^2}  +  \frac{\mathrm{x}_i \cdot \big[ V_{u(i)} \big( V_{u(i)}  -  V_i \mathrm{cos}(\Delta\delta_i)  \big)  \big]}{\mathrm{r}_i^2 + \mathrm{x}_i^2}. $} 
Assuming that $\Delta\delta_i\approx 0$ and $|V_{u(i)}| = |V_{i}|=1 \mathrm{p.u.}$ (under which $\mathrm{sin}(\Delta\delta_i)=\Delta\delta_i$ and $\mathrm{cos}(\Delta\delta_i)=1$) \cite{2},
\begin{subequations}
\begin{align}
P_i + L_i^P  \approx  \frac{\mathrm{r}_i}{\mathrm{r}_i^2 + \mathrm{x}_i^2}    \cdot   \big( V_{u(i)} - V_i \big)         +    \frac{\mathrm{x}_i}{\mathrm{r}_i^2 + \mathrm{x}_i^2} \cdot \big( \Delta\delta_i \big) \label{eq7a}\\ 
Q_i + L_i^Q \approx  \frac{\mathrm{r}_i}{\mathrm{r}_i^2 + \mathrm{x}_i^2} \cdot   \big( \Delta\delta_i \big)        +     \frac{\mathrm{x}_i}{\mathrm{r}_i^2 + \mathrm{x}_i^2} \cdot \big(V_{u(i)} - V_i \big).  \label{eq7b} 
\end{align}
\end{subequations}
Letting, $\Delta V_i = (V_{u(i)} - V_i)$ ,  $Z_i^{\mathrm{r}}=\frac{\mathrm{r}_i}{\mathrm{r}_i^2 + \mathrm{x}_i^2}$ , and $Z_i^{\mathrm{x}}=\frac{\mathrm{x}_i}{\mathrm{r}_i^2 + \mathrm{x}_i^2}$, (7) can be written in terms of $\Delta V_i$ and $\Delta\delta_i$,
\begin{subequations}
\begin{align}
P_i + L_i^P  =  Z_i^{\mathrm{r}} \cdot \Delta V_i + Z_i^{\mathrm{x}} \cdot \Delta \delta_i \label{eq8a}\\ 
Q_i+ L_i^Q  =  Z_i^{\mathrm{x}} \cdot \Delta V_i - Z_i^{\mathrm{r}} \cdot \Delta \delta_i. \label{eq8a}
\end{align}
\end{subequations}
In (8), power flow and loss in line $i$ serving node $i$ is given by the voltage drop $\Delta V_i$ and the voltage angle difference $\Delta \delta_i$  in line $i$. In a more compact and general vector-matrix form,
\begin{equation}
\left[ \begin{array}{c} \mathbf{P} + \mathbf{L}^P \\ \mathbf{Q} + \mathbf{L}^Q \end{array} \right] =\mathbf{Z} \cdot \left[ \begin{array}{c}  \Delta \mathbf{V} \\ \Delta \bm{\delta} \end{array} \right].
\end{equation}
Here, $\mathbf{Z} = \bigl[ \begin{smallmatrix}\mathbf{Z}^\mathrm{r} & \mathbf{Z}^\mathrm{x}\\  \mathbf{Z}^\mathrm{x} &  -\mathbf{Z}^\mathrm{r}\end{smallmatrix}\bigr]$ where $\mathbf{Z}^\mathrm{r} =\Bigg[\begin{smallmatrix} Z_{1}^\mathrm{r} & \dots & 0\\ \vdots & \ddots & \vdots \\ 0 & \dots &  Z_{N}^\mathrm{r}\end{smallmatrix}\Bigg] $ and $\mathbf{Z}^\mathrm{x} =\Bigg[\begin{smallmatrix} Z_{1}^\mathrm{x} & \dots & 0\\ \vdots & \ddots & \vdots \\ 0 & \dots &  Z_{N}^\mathrm{x}\end{smallmatrix}\Bigg] $ are diagonal matrices. Note that, according to the way lines and nodes are indexed, $N$ shows the number of nodes as well as the number of lines in a radial network, hence, $(P,L^P,Q,L^Q,\Delta V_i ,\Delta \delta_i )\in \mathbbm{R}^N$, $\mathbf{Z}\in\mathbbm{R}^{(2N\times2N)}$ and $(\mathbf{Z}^\mathrm{r},\mathbf{Z}^\mathrm{x})\in\mathbbm{R}^{(N\times N)}$. Since $\mathbf{Z}$ is a square invertible matrix, $\Delta \mathbf{V}$ and $\Delta \bm{\delta}$  is therefore given by the linear relation below.
\begin{equation}
\left[ \begin{array}{c} \Delta \mathbf{V} \\ \Delta \bm{\delta}  \end{array} \right] =\mathbf{Z}^{-1} \cdot \left[ \begin{array}{c} \mathbf{P} + \mathbf{L}^P \\ \mathbf{Q} + \mathbf{L}^Q  \end{array} \right].
\end{equation}
The voltage magnitude at node $i$ is given by the substation voltage magnitude minus the total voltage drop from the substation to node $i$ (see Fig. 2), mathematically written as,
\begin{equation}
\mathbf{V}=\mathbf{V}_0 - \bm{\mathcal{U}} \cdot \Delta \mathbf{V}.
\end{equation}
Here, $\bm{\mathcal{U}}\in\mathbbm{R}^{(N\times N)}$ is the upstream node-to-line incidence matrix defined as,
\begin{equation}
[\bm{\mathcal{U}}]_{ij} = \bigg\{\begin{array}{c}1 \qquad \mathrm{if}\,  j\in \bm{\mathcal{U}}(i)\\0 \qquad \mathrm{otherwise.} \end{array}
\end{equation}
where $\bm{\mathcal{U}}(i)$ is the set of lines that connect node $i$ to the substation node. Letting $\mathbf{Z}_V=\mathbf{Z}^{-1} (1:N,1:2N)$ and replacing $\Delta \mathbf{V}$ from (10), (11) can be rewritten as,
\begin{equation}
\mathbf{V}=\mathbf{V}_0 - \bm{\mathcal{U}}\, \mathbf{Z}_V \left[ \begin{array}{c} \mathbf{P} + \mathbf{L}^P \\ \mathbf{Q} + \mathbf{L}^Q  \end{array} \right].
\end{equation}
where $(\mathbf{V},\mathbf{V}_0 )\in \mathbbm{R}^N$ and  $\mathbf{V}_0=[V_0\dots V_0]^T$. In (13), the magnitudes of node voltages are given in terms of line power flows, line losses, and network structure in reference to the substation node voltage. The node voltage magnitude can ultimately be written as a function of power injections $(p_i,q_i)$, line losses $(L_i^P,L_i^Q)$, and other network parameters. Note that $P_i+L_i^P$ and $Q_i+L_i^Q$, as the $i^\mathrm{th}$ element of the RHS of (13), can be written as follows (see Fig. 1),
\begin{subequations}
\begin{align}
&P_i + L_i^P  =  p_i + \sum_{k\in d\{i\}}(P_k + L_k^P) + L_i^P \label{eq14a}\\ 
&Q_i +L_i^Q  =  q_i + \sum_{k\in d\{i\}}(Q_k + L_k^Q) + L_i^Q. \label{eq14a}
\end{align}
\end{subequations}
In vector-matrix form, (14) can be written as,
\begin{subequations}
\begin{align}
&\mathbf{P} +\mathbf{L}^P  = \bm{\mathcal{D}}\,(\mathbf{p}+\mathbf{L}^P) \label{eq15a}\\ 
&\mathbf{Q} +\mathbf{L}^Q  = \bm{\mathcal{D}}\,(\mathbf{q}+\mathbf{L}^Q). \label{eq15b}
\end{align}
\end{subequations}
Here, $\bm{\mathcal{D}}\in \mathbbm{R}^{(N\times N)}$ is the downstream node-to-node (as well as line-to-line) incidence matrix including self (diagonal elements) defined as follows,
\begin{equation}
[\bm{\mathcal{D}}]_{ij} = \bigg\{\begin{array}{cc}1 &\mathrm{if} \, j\in \bm{\mathcal{D}}(i) \, \mathrm{or}\,  i = j \\0 & \mathrm{otherwise.} \end{array}
\end{equation}
Hence, by replacing (15) in (13) and with some modifications,
\begin{equation}
\mathbf{V}=\mathbf{V}_0 - \bm{\mathcal{U}}\, \mathbf{Z}_V \left[ \begin{array}{cc} \bm{\mathcal{D}} &\mathbf{0}\\\mathbf{0} &\bm{\mathcal{D}}\end{array} \right]\left[ \begin{array}{c} \mathbf{p} + \mathbf{L}^P \\ \mathbf{q} + \mathbf{L}^Q  \end{array} \right].
\end{equation}
By letting $\bm{\mathcal{M}}=[\bm{\mathcal{M}}_p\:\bm{\mathcal{M}}_q]=\bm{\mathcal{U}}\, \mathbf{Z}_V \left[ \begin{smallmatrix}\bm{\mathcal{D}} &\mathbf{0}\\\mathbf{0} &\bm{\mathcal{D}}\end{smallmatrix} \right]$, (17) is written,
\begin{equation}
\mathbf{V}=\mathbf{V}_0 - \bm{\mathcal{M}}_p (\mathbf{p}+\mathbf{L}^P)+ \bm{\mathcal{M}}_q (\mathbf{q}+\mathbf{L}^Q).
\end{equation}
where $(\bm{\mathcal{M}}_p,\bm{\mathcal{M}}_q)\in\mathbbm{R}^{(N\times N)}$ are network-related matrices associated with real and reactive power injections and losses. Note that, in reference to the substation voltage $\mathbf{V}_0$, (18) computes the node voltages $\mathbf{V}$, given node injections $(\mathbf{p},\mathbf{q})$, line losses $(\mathbf{L}^P,\mathbf{L}^Q )$, and the network-related matrices $(\bm{\mathcal{M}}_p,\bm{\mathcal{M}}_q)$.  In section IV, (18) is used to set the voltage constraints of the DSO's transactive DAM optimization problem.
\indent The losses $L_i^P$ and $L_i^Q$ are given by the following relations, 
\begin{equation}
\Big\{L_i^P, \,L_i^Q \Big\}  = \bigg\{\mathrm{r}_i \frac {P_i^2 + Q_i^2}{V_i^2}, \,\mathrm{x}_i \frac {P_i^2 + Q_i^2}{V_i^2}\bigg\}.  
\end{equation}
The losses in (19) can be linearized using a first-order Taylor series approximation by omitting the 2nd and higher terms \cite{2},
\begin{subequations}
\begin{align}
\resizebox{1\hsize}{!}{$\mathbf{L}^P(\mathbf{p},\mathbf{q})  \approx \mathbf{L}^P(\mathbf{p^*},\mathbf{q^*}) + \sum_{i} \frac {\partial \mathbf{L}^P}{\partial p_i}(p_i - p_i^*) + \sum_{i} \frac {\partial \mathbf{L}^P}{\partial q_i}(q_i - q_i^*)$}  \label{eq19a}\\ 
\resizebox{1\hsize}{!}{$\mathbf{L}^Q(\mathbf{p},\mathbf{q})  \approx \mathbf{L}^Q(\mathbf{p^*},\mathbf{q^*}) + \sum_{i} \frac {\partial \mathbf{L}^Q}{\partial p_i}(p_i - p_i^*) + \sum_{i} \frac {\partial \mathbf{L}^Q}{\partial q_i}(q_i - q_i^*)$}.   \label{eq19b}
\end{align}
\end{subequations}
In (20), the solution $(\mathbf{p^*},\mathbf{q^*})$ is a feasible power flow solution that is obtained exogenously and set as the center point for the Taylor series approximation. The linear approximations of losses $\mathbf{L}^P$ and $\mathbf{L}^Q$ in (20) are used in (18) and the DSO's DAM optimization problem in section IV. 

\subsection{Probability Efficient Point (PEP) Method}
The inclusion of VRE in the DAM typically entails a considerable deviation between the forecast and the actual production \cite{14}. Due to the stochastic nature of the VRE generation, DAM resource scheduling in a cost-efficient manner to meet the nodal and system demand is a challenging task, especially when the VRE generation does not follow a structured distribution. The PEP method \cite{16,17} can determine VRE generation efficient points at certain confidence levels. The novel and data-driven mathematical program [16] proposed here solves the PEPs using historical VRE generation data without assuming any probability distribution function. The preliminaries of the PEP method are defined below \cite{16,17}. Suppose $\widetilde{\mathbf{p}}\in \mathbbm{R}^{\lvert R\rvert}$ is a stochastic vector, $\mathbf{v}$, $\mathbf{u} \in \mathbbm{R}^{\lvert R\rvert}$ are two realizations, and the probability distribution function $F_{\widetilde{p}} (\widetilde{\mathbf{p}})$ of the stochastic vector $\widetilde{\mathbf{p}}$ is $F_{\widetilde{p}} (\mathbf{v})=\mathrm{Pr}\{\mathbf{v} \geq \widetilde{\mathbf{p}}\}$.\\
$\mathbf{Definition-1:}$ If, $\forall v_m\in\mathbf{v}$ and $\forall u_m\in\mathbf{u}$, $v_m\geq u_m$,  then $\mathbf{v}\geq \mathbf{u}$. Similarly, if $\mathbf{v}\geq \mathbf{u}$, then $v_m\geq u_m$.\\
$\mathbf{Definition-2:}$ Let $\gamma \in (0,1)$. A point $\mathbf{v}^\gamma \in \mathbbm{R}^{|R|}$ is called $\gamma$-efficient point of $F_{\widetilde{p}}$ if $F_{\widetilde{p}} (\mathbf{v}^\gamma) \geq \gamma$ and there is no $\mathbf{u}\leq \mathbf{v}^\gamma$, $\mathbf{u}\neq \mathbf{v}^\gamma$ such that $F_{\widetilde{p}} (\mathbf{u})\geq \gamma$.\\
\indent According to Definition-2, if $\mathbf{v}^\gamma$ is the $\gamma$-efficient point, then $F_{\widetilde{p}} (\mathbf{u}) \geq \gamma$ is equivalent to $\mathbf{u} \geq \mathbf{v}^\gamma$ at a probability level $\gamma$.  That is, a probabilistic constraint $\mathrm{Pr}(\mathbf{u} \geq \widetilde{\mathbf{p}})\geq \gamma$ can be converted to a deterministic constraint $\mathbf{u}\geq \mathbf{v}^\gamma$. The key to this transformation is to calculate the PEP $\mathbf{v}^\gamma$ based on historical samples. Let $\mathbf{\mathcal{S}}$ show the finite set of sample historical realizations of the stochastic vector $\widetilde{\mathbf{p}}=(\widetilde{p}_1,\dots,\widetilde{p}_m,\dots,\widetilde{p}_{|R|})$ for $\lvert R\rvert$ VRE sites in the distribution system. Let $\mathbf{p}^s=(p_1^s,\dots,p_m^s,\dots,p_{\lvert R\rvert}^s)$ show the $s^\mathrm{th}$ sample of the historical realizations of $\widetilde{\mathbf{p}}$, where $s \in S$. Let $\pi^s$ show the probability of each scenario $s$, where $\pi^s=\mathrm{Pr}(\mathbf{p}^s=\widetilde{\mathbf{p}})>0$ and $\sum_{s\in \mathbf{\mathcal{S}}}\pi^s=1$.  According to \cite{16}, the program solving PEP can be formulated as the MILP in (21)-(22).\\
\indent The objective function (21) is the element-wise sum of the PEP vector $\mathbf{v}^\gamma$. Constraint (22a) ensures the total probability of a set of ``selected'' samples is at least equal to $\gamma$. Constraint (22b) guarantees that each element of the solution ($v_m^\gamma \in \mathbf{v}^\gamma$) is higher than the corresponding element of the historical data samples ($p_m^s \in \mathbf{p}^s$). This is ensured through the use of binary variable $\mathbbm{s}^s$ which is equal to unity `1' if all constraints $v_m^\gamma \geq p_m^s, m=1,\dots,\lvert R\rvert$, are met in the $s^{\mathrm{th}}$ sample, and zero otherwise. Note that, assigning the binary variable $\mathbbm{s}^s$ to the whole vector $\mathbf{p}^s$ respects the correlation between various types of VRE in a historical samples.
\begin{equation}
\underset{v_1^\gamma,\dots,v_{\lvert R\rvert}^\gamma}{\text{Minimize}} \: \sum_{m=1}^{|R|}{v_m^\gamma}
\end{equation}
Subject to:
\begin{subequations}
\begin{align}
&\sum_{s \in \mathbf{\mathcal{S}}}{\pi^s\,\mathbbm{s}^s \geq \gamma},\\
&v_m^\gamma \geq p_m^s\,\mathbbm{s}^s,&m=1,\dots,|R|,\:\: s \in S\\
&\mathbbm{s}^s \in \{0,1\}^{\lvert \mathbf{\mathcal{S}}\rvert},\\
&v_m^\gamma \in \mathbbm{R}_{+},&m=1,\dots,\lvert R\rvert
\end{align}
\end{subequations}
The objective function (21) ensures a minimal solution is picked from the selected samples and is referred to as the PEP, $\mathbf{v}^\gamma$, the element-wise sum of which represents the total optimal amount of system VRE generation at probability level $\gamma$ (confidence level $\alpha = 1-\gamma$). By Definition-2, at a confidence level $\alpha$, the VRE at site $m$ is expected to generate an amount, $v_m^\gamma$, that can be replaced with $\tilde{p}_{i,t}^R$ in the power balance equation, (1), to compute the DLMP values. It is worth mentioning that, the proposed PEP method does not guarantee a unique solution. However, given the randomness of the historical samples, as is the case in our simulation, the nonuniqueness may rarely become an issue. Nevertheless, if it becomes an issue, one can find multiple PEP solutions and use a scenario-based stochastic optimization method to solve this problem \cite{wu2017stochastic}.
\subsection{DER with Temporal Constraints and Bidding Framework}
DERs such as BESSs require additional temporal constraints that must be satisfied during DAM scheduling. In this section, we summarize the set of constraints, $\mathbbm{B}(x_t )$, that consider the most necessary aspects of optimally operating a BESS. With $x_t$ showing any variable at time $t$, $\mathbbm{B}(x_t )$ is listed in (23).\\
\indent The BESS unit's state of charge (SOC) is given by (23a). A small amount of its energy is dissipated due to self-discharge (``leakage"), or during charging and discharging. These effects are respectively accounted for using the scaling coefficients, $\beta_{i,t}$, $\widecheck{\beta}_{i,t}$, and $\widehat{\beta}_{i,t}$, each of which lie in the interval $(0,1]$. The BESS unit's SOC has to remain within SOC limits (23b). The difference in extraction $\widehat{p}_{i,t}^B$ and injection $\widecheck{p}_{i,t}^B$ is modeled as the net grid extraction $p_{i,t}^B$ (23c). Both $\widecheck{p}_{i,t}^B$ and $\widehat{p}_{i,t}^B$ are bound by the BESS charging and discharging power rates when the unit is `ON' and scheduled for charging (23d) or dishcarging (23e). Other integer constraints, such as minimum number of consecutive charging/discharging hours, etc., are summarized in (23f) (see our work in \cite{13} for more details).
\begin{subequations}
\begin{align}
&e_{i,t}^B=\beta_{i,t}\,e_{i,t-1}^B+\Delta t (\widehat{\beta}_{i,t}\,\widehat{p}_{i,t-1}^B - \frac{\widecheck{p}_{i,t-1}^B}{\widecheck{\beta}_{i,t}})\\
&\underline{\mathcal{E}}_i \leq e_{i,t}^B \leq \overline{\mathcal{E}}_i\\
&p_{i,t}^B = \widehat{p}_{i,t}^B - \widecheck{p}_{i,t}^B\\
&-\underline{P}_i^B \, \widecheck{\mathbbm{z}}_{i,t} \leq  \widecheck{p}_{i,t}^B \leq \overline{P}_i^B \, \widecheck{\mathbbm{z}}_{i,t}\\
&-\underline{P}_i^B \, \widehat{\mathbbm{z}}_{i,t} \leq  \widehat{p}_{i,t}^B \leq \overline{P}_i^B \, \widehat{\mathbbm{z}}_{i,t}\\
&\mathbbm{M}_{i,t} \leq 0
\end{align}
\end{subequations}
\indent Similar to the wholesale market, participants place power and monetary amount bids and offers in segments. A seller participant, such as a CG at node $i$, offers $\{c_{i,1,t}^G,\dots ,c_{i,w,t}^G \}$, $\{p_{i,1,t}^G,\dots,p_{i,w,t}^G \}$ at timeslot $t$. Likewise, a buyer participant, such as an RL, places bid and amount segments as  $\{c_{i,1,t}^D,\dots, c_{i,w,t}^D \}$, $\{p_{i,1,t}^D,\dots, p_{i,w,t}^D \}$ at timeslot $t$. In light of FERC's recent rule on Electric Storage Participation in Markets \cite{20}, we assume that BESS units participate by bidding injection and extraction prices. Energy is traded at the LMP with the wholesale and the VREs place zero-price offers in accordance with current practices in the market.

\section{Problem Formulation}
\subsection{DAM Optimization Problem}
The DSO's transactive DAM optimization problem can be formulated as follows:\\
\begin{equation}
\underset{x_t}{\text{Minimize}} \quad \mathcal{O}_p(x_t)+\mathcal{O}_q(x_t)\qquad 
\end{equation}
Subject to:
\begin{subequations}
\begin{align}
&\resizebox{0.55\hsize}{!}{$p_0(x_t)+\sum_{i}{\big(p_i(x_t)} - L_i^P(x_t)\big)=0$}\\
&\resizebox{0.55\hsize}{!}{$q_0(x_t)+\sum_{i}{\big(q_i(x_t)} - L_i^Q(x_t)\big)=0$}\\
&p_i(x_t)=p_i^D(x_t)-p_i^G(x_t)-p_i^B(x_t)-v_{i,t}^\gamma\\
&q_i(x_t)=q_i^D(x_t)-q_i^G(x_t)-q_i^B(x_t)-\kappa_i^R v_{i,t}^\gamma\\
&1-\epsilon \leq V(x_t) \leq 1+\epsilon \\
&\lvert P_i(x_t)\rvert + \lvert Q_i(x_t) \rvert \leq \sqrt{2}\,\overline{S}_i \\
&x_t \in \Big\{\mathcal{G}(x_t)\cup \mathcal{D}(x_t) \cup \mathcal{R}(x_t) \cup \mathbbm{B}(x_t )\Big\}
\end{align}
\end{subequations}
where $x_t \in \big\{p_{i,w,t}^G,p_{i,w,t}^D, p_{0,t},\widecheck{p}_{i,t}^B,\widehat{p}_{i,t}^B, q_{i,w,t}^G, q_{0,t},q_{i,w,t}^D\big\}$,
\begin{equation*}
\begin{split}
\mathcal{O}_p(x_t)=&\sum_{i,w,t}{c_{i,w,t}^{G,p}\cdot p_{i,w,t}^G}+\lambda_p^0 \cdot p_{0,t} + \sum_{i,t}{\widehat{c}_{i,t}^{B,p} \cdot \widehat{p}_{i,t}^B }\\
- &\sum_{i,t}{\widecheck{c}_{i,t}^{B,p} \cdot \widecheck{p}_{i,t}^B } -\sum_{i,w>1,t}{c_{i,w,t}^{D,p}\cdot p_{i,w,t}^D} + \sum_{i,w,t}{\zeta_{i,t}  \cdot y_{i,w,t}}\\ 
\mathcal{O}_q(x_t)=&\sum_{i,w,t}{c_{i,w,t}^{G,q}\cdot q_{i,w,t}^G}+\lambda_p^0 \cdot q_{0,t} -\sum_{i,w>1,t}{c_{i,w,t}^{D,q}\cdot q_{i,w,t}^D }
\end{split}
\end{equation*}
\indent The above optimization problem minimizes the objective function in (24) by scheduling least cost generation and wholesale supply, least cost energy extraction, and high value energy injections and loads while reducing curtailments and meeting grid and DER constraints in (25). The fifth term in $\mathcal{O}_p(x_t)$ is modeled for $w>1$ to account only for demand segments that place bids; we assume the first segment of the demand as must-serve loads. The sixth term in $\mathcal{O}_p(x_t)$ accounts for curtailments to achieve feasibility. Constraints (25a)-(25d) pertain to the system and nodal power-balance-equations. Note in (25a) and (25b) that the losses, $L_i^P(x_t)$ and $L_i^Q(x_t)$, are given by the linearized constraints in (20). Constraints (25e) and (25f) enforce voltage limit bounds and the line limit constraints respectively. To obtain congestion prices, the quadratic line limit constraints $[P_i (x_t)]^2+[Q_i (x_t)]^2 \leq \overline{S}_i^2$ are linearized as in (25f) using an outer approximation \cite{18}.
\subsection{DLMP Decomposition}
In (26), real and reactive power DLMPs, $\Omega_{p,i}$ and $\Omega_{q,i}$, at node $i$ are defined as the sum of DLMP components, $\Omega_{p,i}^E, \Omega_{p,i}^L,\Omega_{p,i}^V, \Omega_{p,i}^C$ and $\Omega_{q,i}^E, \Omega_{q,i}^L,\Omega_{q,i}^V, \Omega_{q,i}^C$ corresponding to energy, loss, voltage violation, and congestion prices.\\
\begin{subequations}
\begin{align}
&\Omega_{p,i} =\Omega_{p,i}^E+\Omega_{p,i}^L+\Omega_{p,i}^V+\Omega_{p,i}^C \nonumber\\
&=\big(\Omega_{p,i}^E\big)+\bigg( \Omega_{p,i}^E\cdot \sum_{j}{\frac{\partial L_j^P}{\partial p_i}}+\Omega_{q,i}^E\cdot \sum_{j}{\frac{\partial L_j^Q}{\partial p_i}}\bigg) \nonumber\\ 
&+ \bigg( \sum_{i'}{\big(\mu_{i'}^\mathrm{min} - \mu_{i'}^\mathrm{max}\big)\frac{\partial V_{i'}}{\partial p_i}}\bigg) +  \bigg( \sum_{j\in u\{i\}}{\rho_j\frac{\partial S_{j}}{\partial p_i}}\bigg)\\
&\Omega_{q,i} =\Omega_{q,i}^E+\Omega_{q,i}^L+\Omega_{q,i}^V+\Omega_{q,i}^C \nonumber\\
&=\big(\Omega_{q,i}^E\big)+\bigg(\Omega_{q,i}^E\cdot \sum_{j}{\frac{\partial L_j^Q}{\partial q_i}}+\Omega_{p,i}^E\cdot \sum_{j}{\frac{\partial L_j^P}{\partial q_i}}\bigg) \nonumber \\
&+ \bigg( \sum_{i'}{\big(\mu_{i'}^\mathrm{min} - \mu_{i'}^\mathrm{max}\big)\frac{\partial V_{i'}}{\partial q_i}}\bigg) +  \bigg( \sum_{j\in u\{i\}}{\rho_j\frac{\partial S_{j}}{\partial q_i}}\bigg)
\end{align}
\end{subequations}
\begin{subequations}
\begin{align}
\frac{\partial L_j^P}{\partial p_i}=\bigg(2P_j\frac{\partial P_j}{\partial p_i} +2Q_j\frac{\partial Q_j}{\partial p_i}\bigg)\cdot \frac{\mathrm{r}_j}{V_i^2} \\
\frac{\partial L_j^P}{\partial q_i}=\bigg(2P_j\frac{\partial P_j}{\partial q_i} +2Q_j\frac{\partial Q_j}{\partial q_i}\bigg)\cdot \frac{\mathrm{r}_j}{V_i^2} \\
\frac{\partial L_j^Q}{\partial p_i}=\bigg(2P_j\frac{\partial P_j}{\partial p_i} +2Q_j\frac{\partial Q_j}{\partial p_i}\bigg)\cdot \frac{\mathrm{x}_j}{V_i^2} \\
\frac{\partial L_j^Q}{\partial q_i}=\bigg(2P_j\frac{\partial P_j}{\partial q_i} +2Q_j\frac{\partial Q_j}{\partial q_i}\bigg)\cdot \frac{\mathrm{x}_j}{V_i^2} 
\end{align}
\end{subequations}
\begin{subequations}
\begin{align}
&\frac{\partial P_j}{\partial p_i}=\bm{\mathcal{D}}(j,i) + \sum_{k\in d(j)}{\bm{\mathcal{D}}(j,k)\frac{\partial L_k^P}{\partial p_i}} \\
&\frac{\partial P_j}{\partial q_i}=\sum_{k\in d(j)}{\bm{\mathcal{D}}(j,k)\frac{\partial L_k^P}{\partial q_i}} \\
&\frac{\partial Q_j}{\partial p_i}=\sum_{k\in d(j)}{\bm{\mathcal{D}}(j,k)\frac{\partial L_k^Q}{\partial p_i}} \\
&\frac{\partial Q_j}{\partial q_i}=\bm{\mathcal{D}}(j,i) + \sum_{k\in d(j)}{\bm{\mathcal{D}}(j,k)\frac{\partial L_k^Q}{\partial q_i}}  
\end{align}
\end{subequations}
\indent The energy component of the DLMPs, $\Omega_p^E, \Omega_q^E$, are given by the shadow prices of (25a) and (25b). These prices show the marginal unit price that are typically, but not always, equal to the LMP $(\lambda_p^0,\lambda_q^0)$ at the substation node.\\
\indent The loss component of real and reactive power prices, $\Omega_{p,i}^L$ and $\Omega_{q,i}^L$, can be computed as follow. Assuming constant voltage at node $i$ in (19), the sensitivity of real and reactive power losses on line $j$ with respect to real and reactive power injections at node $i$ can be given by (27). Further simplifications using (14) yields the line flow sensitivities in (28). By replacing (28) in (27) and solving recursively, the loss sensitivities $\sum_{j}{\frac{\partial L_j^P}{\partial p_i}}$, $\sum_{j}{\frac{\partial L_j^P}{\partial q_i}}$, $\sum_{j}{\frac{\partial L_j^Q}{\partial p_i}}$, and $\sum_{j}{\frac{\partial L_j^Q}{\partial q_i}}$ can be computed to obtain $\Omega_{p,i}^L$ and $\Omega_{q,i}^L$, in (26).\\
\indent The voltage violation prices, $\Omega_{p,i}^V$ and $\Omega_{q,i}^V$, can be similarly derived by finding $\frac{\partial V_i}{\partial p_i'}$ and $\frac{\partial V_i}{\partial q_i'}$ according to (18) as,
\begin{subequations}
\begin{align}
\frac{\partial V_i}{\partial p_i'}=-\bm{\mathcal{M}}_p(i,i')-\sum_{k=1}^{N}{\bm{\mathcal{M}}_p(i,k)\cdot\frac{\partial L_k^P}{\partial p_i'}}\nonumber\\
- \sum_{k=1}^{N}{\bm{\mathcal{M}}_q(i,k) \cdot \frac{\partial L_k^Q}{\partial p_i'}}\\
\frac{\partial V_i}{\partial q_i'}=-\bm{\mathcal{M}}_q(i,i')-\sum_{k=1}^{N}{\bm{\mathcal{M}}_q(i,k)\cdot \frac{\partial L_k^Q}{\partial q_i'}}\nonumber\\
- \sum_{k=1}^{N}{\bm{\mathcal{M}}_p(i,k) \cdot \frac{\partial L_k^P}{\partial q_i'}}
\end{align}
\end{subequations}
Hence, by using (29) and the shadow prices of (25e), $\mu_{j}^\mathrm{min}$, $\mu_{j}^\mathrm{max}$, the voltage violation components, $\Omega_{p,i}^V$, $\Omega_{q,i}^V$, in (26) can be readily computed.\\
\indent The congestion component prices, $\Omega_{p,i}^C$, $\Omega_{q,i}^C$, defined in (26) are obtained by expanding (25f) into four separate constraints $\pm P_i\pm Q_i \leq \sqrt{2}\,\overline{S}_i$ with shadow prices $\{\rho_{++}, \rho_{--},\rho_{+-},\rho_{-+}\}$.  By letting $\rho_{(p,q),1}=\rho_{++}-\rho_{--}+\rho_{+-}-\rho_{-+}$, and $\rho_{(p,q),2}=\rho_{++}-\rho_{--}-\rho_{+-}+\rho_{-+}$, the congestion DLMP component can be written as,
\begin{subequations}
\begin{align}
\sum_{j\in u\{i\}}{\rho_j\frac{\partial S_j}{\partial p_i}}=\sum_{j=1}^{N}{\bm{\mathcal{U}}(i,j) \cdot \rho_{p,1}(j) \cdot \frac{\partial P_j}{\partial p_i}}\nonumber\\ 
+\sum_{j=1}^{N}{\bm{\mathcal{U}}(i,j) \cdot \rho_{p,2}(j) \cdot \frac{\partial Q_j}{\partial p_i}}\\
\sum_{j\in u\{i\}}{\rho_j\frac{\partial S_j}{\partial q_i}}=\sum_{j=1}^{N}{\bm{\mathcal{U}}(i,j) \cdot \rho_{q,2}(j) \cdot \frac{\partial Q_j}{\partial q_i}}\nonumber\\
+\sum_{j=1}^{N}{\bm{\mathcal{U}}(i,j) \cdot \rho_{p,1}(j) \cdot \frac{\partial P_j}{\partial q_i}}
\end{align}
\end{subequations}

\noindent By using (28) in (30), the real and reactive power congestion prices, $\Omega_{p,i}^C$  and  $\Omega_{q,i}^C$, in (26) can be computed.

\section{Simulation Results}
GAMS/CPLEX was used to program and simulate the proposed MILP models on a modified IEEE 69-node system as shown in Fig. 2. The total load demand in each node was split into three segments of 50\%, 25\%, and 25\%. The first load segments were considered fixed without any bids. The second and third segments were assumed price-responsive with associated bid amounts randomly generated in the range [13.8, 28.1] and [10.3, 26.5] \$/MWhr for real power and 30\% of these amounts for reactive power. The VREs were assumed to bid 0 \$/MWhr and the BESS was considered to sell at extraction price of 25 \$/MWhr and buy at injection price of 20 \$/MWhr. An hourly LMP of a typical-day varying in the range [16.45, 42.14] \$/MWhr was considered at the substation node. Western Wind and Solar Integration Study (WWSIS) historical data set \cite{21} for Wind and PV generation at four different sites were used to implement the PEP method and obtain the VRE generation at different confidence levels and incorporate them into the DSO's transactive DAM model. VRE units located at close proximity were considered identical and assigned the same set of historical data. A sample wind and solar daily generation profiles of the WWSIS for 2013 along with the computed PEPs, $v^\alpha$, at confidence levels $\alpha$ = 0.25 and 0.75 are shown in Fig. 3. As expected, with more $\alpha$, the PEP method returns more generation, as determined by the historical data samples ($\mathbf{p}^s$) and the probability of their occurence ($\pi^s$). Note that, in this study we assume the DSO knows these probabilities and the confidence levels a priori. The VRE generation of additional sites can be thought as additional layers to Fig. 3--coupled by the binary variable $\mathbbm{s}^s$ at each timeslot. Hence, the DSO runs the data-driven PEP method to determine the efficient points for the minimal VRE generation in order to incorporate them into the DAM.
\begin{figure}[!htb]
\begin{center}
\includegraphics[width=3.5in]{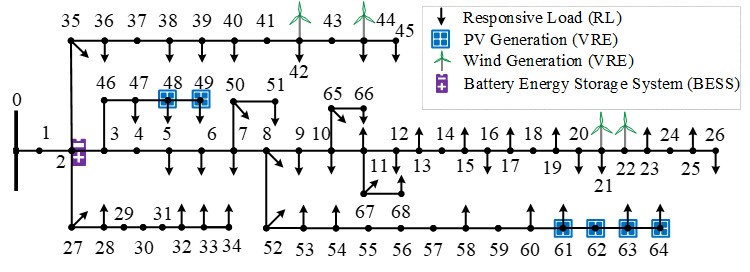}
\caption{Modified IEEE 69-node system with diverse DER.}
\label{fig2}
\end{center}
\end{figure}
\setlength{\textfloatsep}{5pt plus 1.0pt minus 2.0pt}


\indent Four Scenarios (I-IV), with 200\% system load demand in Scenarios I and II and 100\% in Scenarios III and IV, were simulated to show the effect of the VRE penetrations on real and reactive power DLMPs and their components. In Scenario I, no VRE penetration was considered. With $K$ being a scaling of the VRE output, 50 ($K$=1), 100 ($K$=2), and 200\% ($K$=4) VRE output penetrations at a confidence level of $\alpha$ = 0.75 were considered in scenarios II-IV. Fig. 4 shows these results (left column for $\Omega_{p}$, right column for $\Omega_{q}$) at node 60 where the majority of the system load is located. Here, the four DLMP components are shown by bars. The algebraic sum of the components (the DLMP) is shown by the red graph with asterisks. The substation LMP is also added as the green line graph with star markers. The blue bar graph shows the energy component ($\Omega_{p}^E$, $\Omega_{q}^E$) of the DLMP as the marginal price of energy given by the Lagrange multipliers of (25a-b). Typically, $\Omega_{p}^E$ and $\Omega_{q}^E$ are equal to the real and reactive LMP ($\lambda_p^0, \lambda_q^0$) at the substation node if it serves the next unit of energy (MWhr, MVARhr) as the marginal node. We define MVARhr as the amount of MVARs served in an hour.\\
\begin{figure}[!htb]
\begin{center}
\includegraphics[width=3.5in]{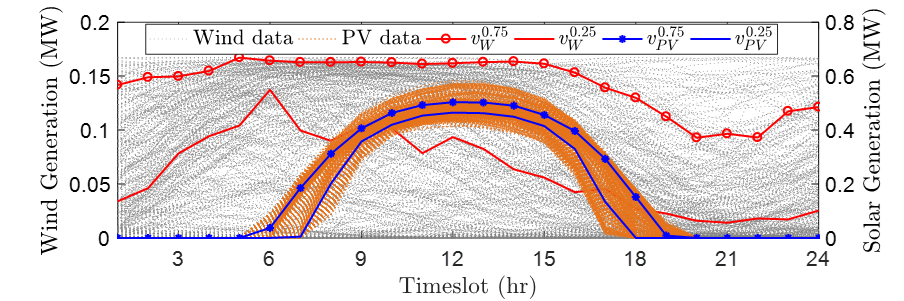}
\caption{Sample historical wind (W) and PV generation daily profiles and the PEP outcome at 25\% (no marker) and 75\% (with marker) confidence levels.}
\label{fig4}
\end{center}
\end{figure}
\setlength{\textfloatsep}{5pt plus 1.0pt minus 2.0pt}
\indent In Scenario I, due to high demand, a voltage violation at node 64 and a congestion at line 2 causes the voltage and congestion components of the DLMP ($\Omega^V_{p,60}$, $\Omega^C_{p,60}$ and $\Omega^V_{q,60}$, $\Omega^C_{q,60}$) to appear during timeslots 16-18. These values show the per unit price incurred to node 60 due to the participation of its consumption in increasing the voltage violations and power flow congestions of other nodes and lines in the system. The loss components, $\Omega^L_{p,60}$, $\Omega^L_{q,60}$, are inherent and appear in all timeslots as losses are unavoidable, i.e., injection of any amount at a node incurs an inevitable transmission loss cost. The marginal node cost shows the energy price component ($\Omega^E_{p}$, $\Omega^E_{q}$) given by the equilibrium point of the supply-demand curve, which in this scenario happens to be the substation node as the blue bar graphs ($\Omega^E_{p,60}$, $\Omega^E_{q,60}$) coincide with the green line graphs with star markers ($\lambda^0_p$, $\lambda^0_p$).   \\ 
\indent By adding the VRE units (as in Fig. 2) in Scenario II  ($K$=1 and $\alpha$=0.75), the congestion prices, $\Omega^C_{p,60}$ and $\Omega^C_{q,60}$, and the voltage violation prices, $\Omega^V_{p,60}$ and $\Omega^V_{q,60}$, during hours 16-18 completely vanish. This is because the deployment of the zero marginal cost VRE units reinforce the grid by reducing the power flow burden on line 1 and other lines, and by serving the high demand at node 60 from nearby nodes (61-64), causing lesser voltage drops and hence removing the congestion and voltage violation prices. Notice that during timeslots 10-12, the LMP ($\lambda_p^0$) does not coincide with the energy price ($\Omega_{p,60}^E$). This shows that the system's real power demand has been entirely satisfied by the VRE units' generation. During these hours, no real power is drawn from the substation node and a load segment bid lower than $\lambda_p^0$ has determined the energy equilibrium price as the marginal unit. Note, however, that this is not the case for $\Omega_{q,60}^E$ and $\lambda^0_q$ as we have set the VREs to allow only up to 30\% of their real power generation as their reactive power output using inverters. Hence, the lack of reactive power supply in the system requires MVAR injections from the substation node resulting in equal $\lambda^0_q$ and $\Omega^E_{q,60}$. 
\begin{figure}[!htb]
\begin{center}
\includegraphics[width=3.5in]{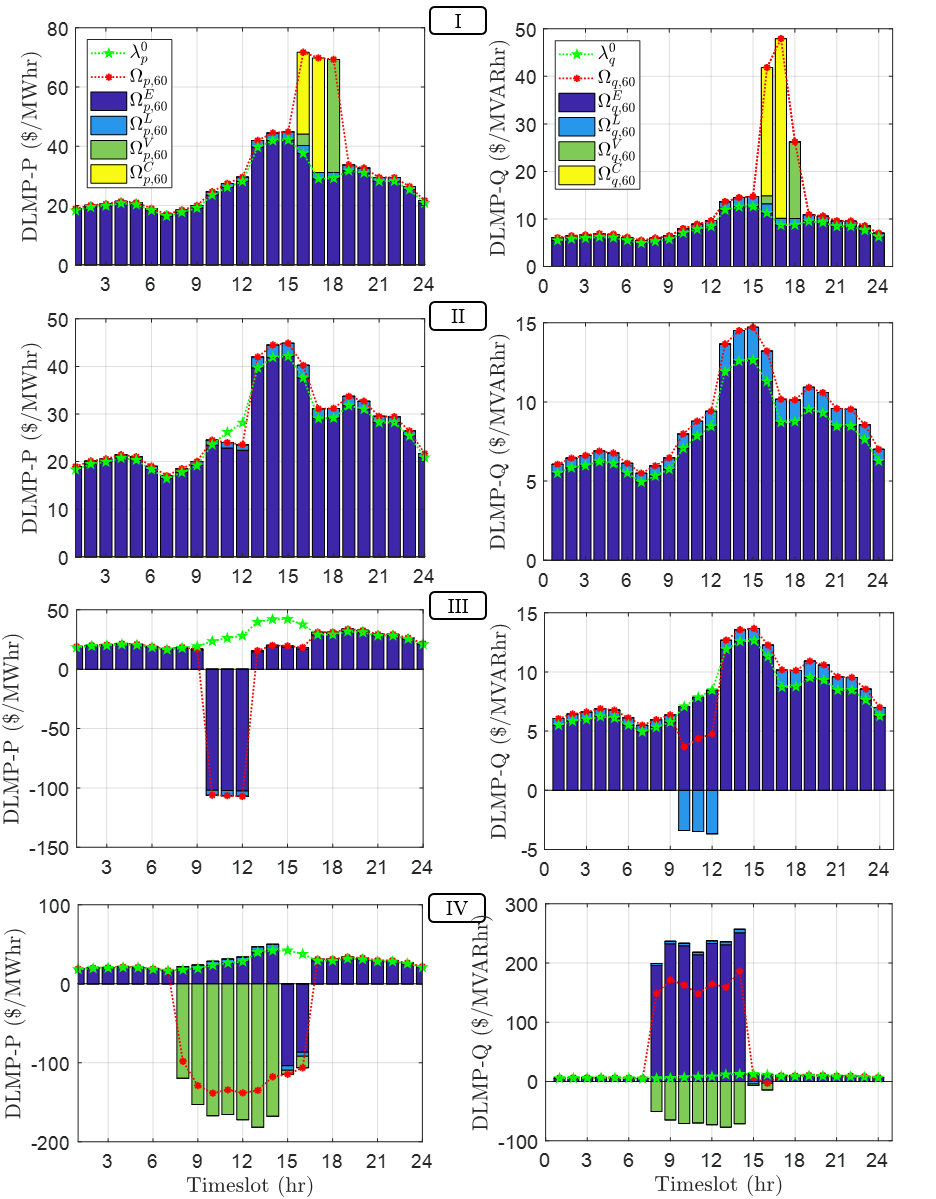}
\caption{Real and reactive DLMPs  (DLMP-P, DLMP-Q) of node 60 ($\Omega_{p,60}$, $\Omega_{q,60}$), their components (energy ($\Omega_{p}^E$, $\Omega_{q}^E$), loss ($\Omega_{p}^L$, $\Omega_{q}^L$), voltage violation ($\Omega_{p}^V$, $\Omega_{q}^V$), and congestion ($\Omega_{p}^C$, $\Omega_{q}^C$)), and LMPs ($\lambda^0_{p}$, $\lambda^0_{q}$) in Scenarios I-IV.}
\label{fig5}
\end{center}
\end{figure} 
\setlength{\textfloatsep}{5pt plus 1.0pt minus 2.0pt}

\indent In Scenario III, with more VRE penetration ($K$=2 and $\alpha$ = 0.75), no violations are seen. However, negative DLMP and loss components are seen. Due to VRE generation curtailment cost of \$100/MWhr, a negative $\Omega^E_{p,60}$  and $\Omega^L_{p,60}$ are seen during timeslots 10-12 where VRE is curtailed to achieve a feasible solution. In other words, node 60 is being paid at the value of lost VRE generation for consuming real power and for causing losses ($\Omega^L_{p,60}$= -\$4.085 /MWhr). On the other hand, $\Omega^E_{q,60}$ and $\Omega^L_{q,60}$ are positive during all timeslots except 10-12 for the loss component. This is because the system demand for reactive power is high and VREs do not curtail reactive power. The reactive power generation of VREs are based on its inverter's reactive power generation capability (set to 0-30\% of their real power generation for this study). As a result, during timeslots 10-12, the reactive power consumption at node 60 is being rewarded at \$3.40/MVARhr for the loss component as it increases real power losses (27b) resulting in lesser VRE curtailment and lower curtailment costs. Such intricate details in reactive power consumption and component-wise pricing is quite interesting, which to the best of our knowledge, has not been studied in previous work.\\
\indent In Scenario IV ($K$=4 and $\alpha$=0.75), due to high VRE penetration, high reverse power flows are seen causing several nodes to violate the upper voltage limit. These effects are seen as negative  $\Omega^V_{p,60}$ and $\Omega^V_{q,60}$. As the largest sink, node 60 is rewarded during hours 8-14 for its consumption--maintaining a lower system voltage and avoiding further VRE curtailments during timeslots 15-16. However, the consumption of reactive power comes at a high price despite the reward of $\Omega^V_{q,60}$. These results indicate that while at times there maybe negative $\Omega_{p}$, it might be unlikely to have negative $\Omega_{q}$. Depending on the system's reactive power supply and demand, the net DLMP ($\Omega_{p}+\Omega_{q}$) could be positive or negative, even if $\Omega_{p}$ is negative--underlining the importance of reactive power pricing.\\
\indent The effect of PEP on $\Omega_{p}$ and  $\Omega_{q}$ of node 60 at different confidence levels with low (K=1) and high (K=4) VRE penetration is shown in Fig. 5. As expected, under both low (solid lines) and high (dotted lines) VRE penetration, $\Omega_{p}$,  $\Omega_{q}$ are mostly affected by the confidence levels during mid-day hours when there is high solar generation. At low penetration, while increasing confidence levels decreases $\Omega_{p}$, it has no effect on $\Omega_{q}$. At high penetration, high confidence levels do not necessarily entail higher negative $\Omega_{p}$ or $\Omega_{q}$. Network constraint violations, mainly due to voltage and congestion, as well as bidding and historical data samples in the PEP method are the main factors that determine $\Omega_{p}$ and  $\Omega_{q}$. \\
\begin{figure}[!htb]
\begin{center}
\includegraphics[width=3.5in]{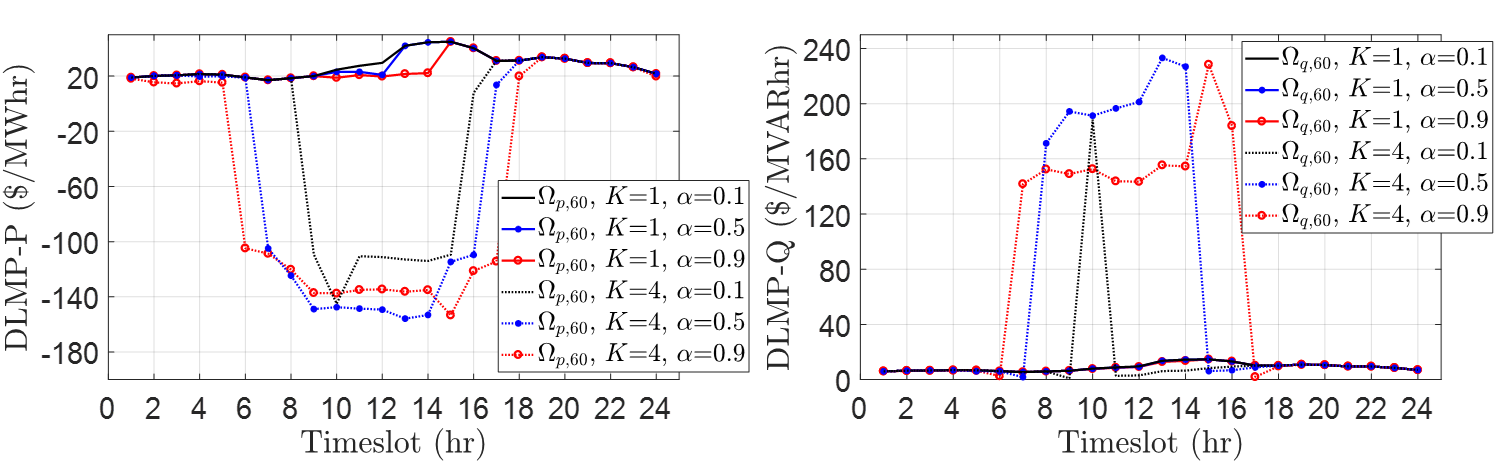}
\caption{Real, reactive DLMPs ($\Omega_{p,60}$, $\Omega_{q,60}$) at different confidence levels.}
\label{fig6}
\end{center}
\end{figure}
\setlength{\textfloatsep}{5pt plus 1.0pt minus 2.0pt}

\indent Lastly, to show the effect of temporally-constrained DER on $\Omega_{p}$ and  $\Omega_{q}$, Scenarios I and III were simulated with a BESS located at node 2. Fig. 6 (left) shows the BESS's real power schedule ($P_2^B$), injection ($\check{c}^{B,p}_2$) and extraction ($\hat{c}^{B,p}_2$) prices, and $\Omega_{p,2}$ ($\Omega_{q,2}$ is skipped for clarity). Notice that, $P_2^B$ follows the DLMP peak and valley. It charges when $\Omega_{p,2}$ is lower than $\check{c}^{B,p}_2$ and discharges when it is higher than $\hat{c}^{B,p}_2$. Fig. 6 (right) shows DLMPs before ($\Omega_{p,2}$, $\Omega_{q,2}$) and after ($\Omega_{p,2}^B$, $\Omega_{q,2}^B$) adding the BESS in Scenario III. As seen, the BESS charges during hours of highly negative $\Omega_{p,2}$ (preventing VRE curtailments) and discharges during peak $\Omega_{p,2}$. Note that, the BESS does not influence $\Omega_{q,2}$ as it does not generate reactive power. These results suggest that reactive power pricing should be treated separately from its real power counterpart.\\ 
\begin{figure}[!htb]
\begin{center}
\includegraphics[width=3.5in]{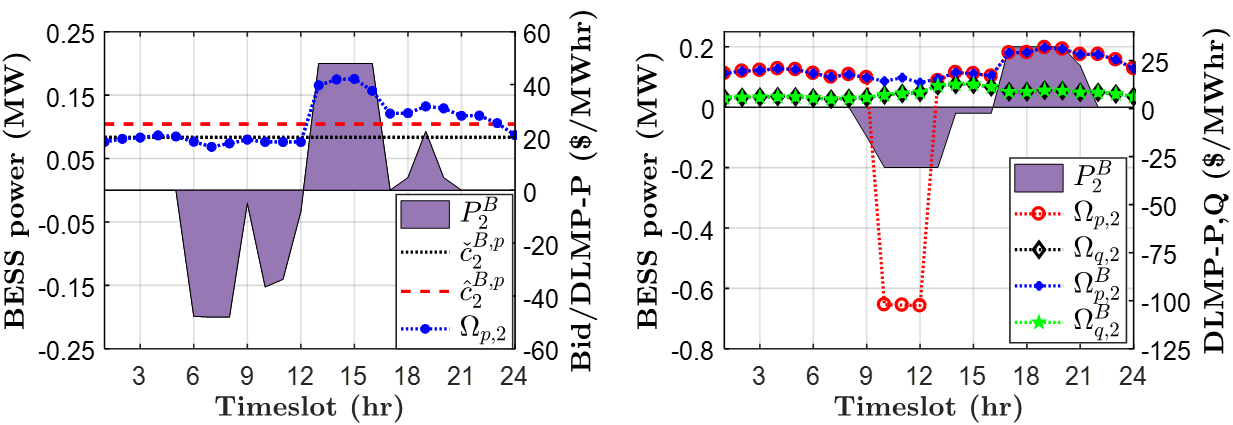}
\caption{BESS charging and discharging schedule versus its bid and DLMP-P.}
\label{fig7}
\end{center}
\end{figure}
\setlength{\textfloatsep}{2pt plus 1.0pt minus 2.0pt}
\section{Conclusions}
A DLMP-based DAM that efficiently accounts for various penetration scales of VRE while accounting for its uncertainty through the data-driven PEP mehtod has been proposed. DLMP components for energy, losses, voltage violation, and congestion were derived through comprehensive sensitivity analyses using reasonable approximations. The effect of VRE penetrations, BESSs, and network constraints on DLMP components under forward and reverse flow were explained. The significance of reactive power pricing was demonstrated. The proposed approach can be easily extended to unbalanced distribution systems by adding a coupling constraint across phases and specifying the maximum allowable power imbalance. Additionally, forecasting tools could be leveraged and incorporated within the PEP method to add further efficiency \cite{17}. Other aspects, such as sharing mechanisms to (re)distribute DERs' investment costs, revenues, etc., which upon allocation mitigate voltage violations and congestions to several nodes and lines, are further interesting research directions. 
\ifCLASSOPTIONcaptionsoff
  \newpage
\fi

\includepdf[page={1}]{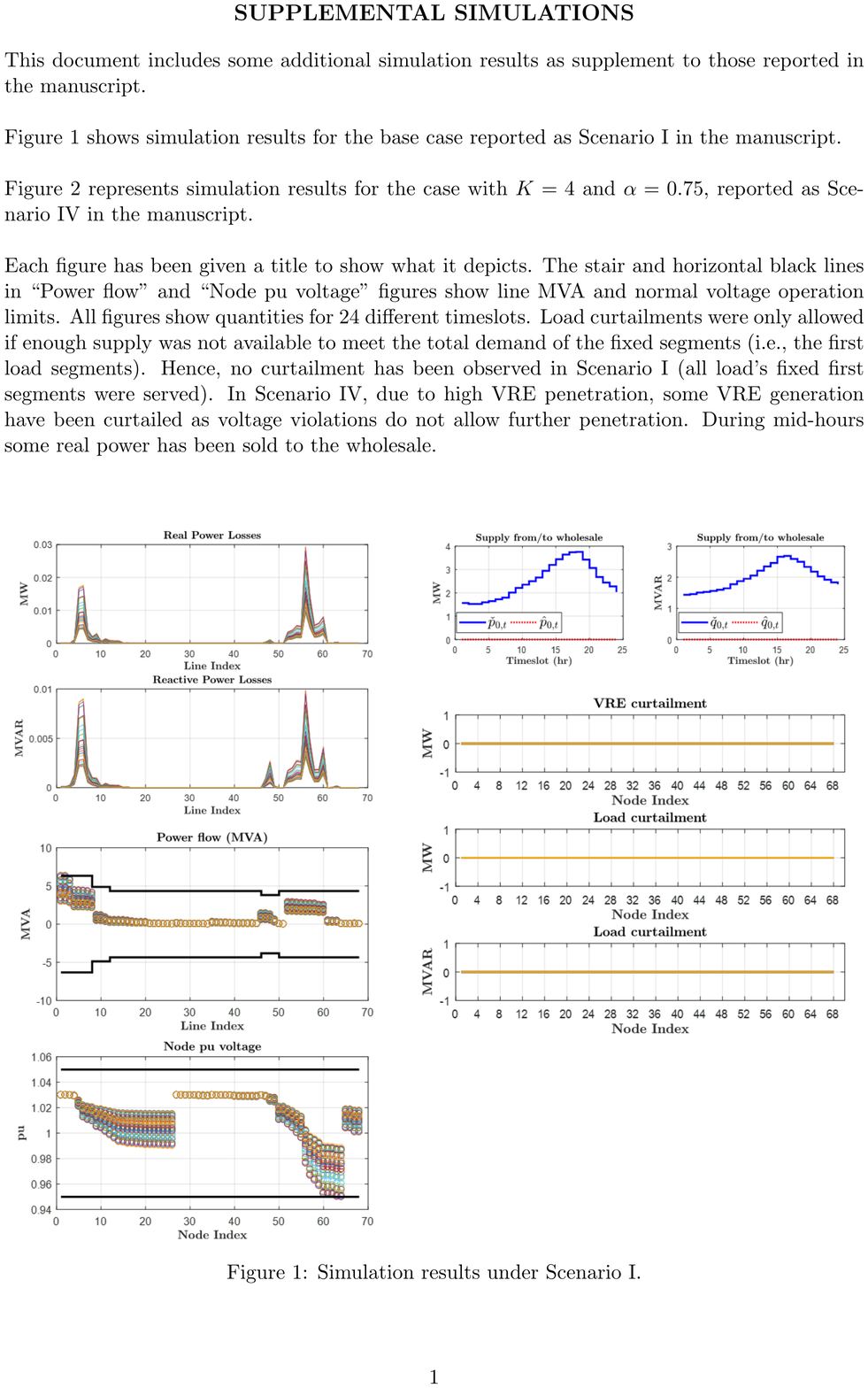} 
\includepdf[page={2}]{SUPPLEMENTAL} 

\begin{thebibliography}{10}
\providecommand{\url}[1]{#1}
\csname url@samestyle\endcsname
\providecommand{\newblock}{\relax}
\providecommand{\bibinfo}[2]{#2}
\providecommand{\BIBentrySTDinterwordspacing}{\spaceskip=0pt\relax}
\providecommand{\BIBentryALTinterwordstretchfactor}{4}
\providecommand{\BIBentryALTinterwordspacing}{\spaceskip=\fontdimen2\font plus
\BIBentryALTinterwordstretchfactor\fontdimen3\font minus
  \fontdimen4\font\relax}
\providecommand{\BIBforeignlanguage}[2]{{%
\expandafter\ifx\csname l@#1\endcsname\relax
\typeout{** WARNING: IEEEtran.bst: No hyphenation pattern has been}%
\typeout{** loaded for the language `#1'. Using the pattern for}%
\typeout{** the default language instead.}%
\else
\language=\csname l@#1\endcsname
\fi
#2}}
\providecommand{\BIBdecl}{\relax}
\BIBdecl

\bibitem{1}
S.~Huang, Q.~Wu, S.~S. Oren, R.~Li, and Z.~Liu, ``Distribution locational
  marginal pricing through quadratic programming for congestion management in
  distribution networks,'' \emph{IEEE Transactions on Power Systems}, vol.~30,
  no.~4, pp. 2170--2178, 2015.

\bibitem{2}
H.~Yuan, F.~Li, Y.~Wei, and J.~Zhu, ``Novel linearized power flow and
  linearized opf models for active distribution networks with application in
  distribution lmp,'' \emph{IEEE Transactions on Smart Grid}, 2016.

\bibitem{3}
L.~Bai, J.~Wang, C.~Wang, C.~Chen, and F.~F. Li, ``Distribution locational
  marginal pricing (dlmp) for congestion management and voltage support,''
  \emph{IEEE Transactions on Power Systems}, 2017.

\bibitem{4}
S.~Hanif, K.~Zhang, C.~Hackl, M.~Barati, H.~B. Gooi, and T.~Hamacher,
  ``Decomposition and equilibrium achieving distribution lmp using trust-region
  method,'' \emph{IEEE Transactions on Smart Grid}, 2018.

\bibitem{5}
M.~Caramanis, E.~Ntakou, W.~W. Hogan, A.~Chakrabortty, and J.~Schoene,
  ``Co-optimization of power and reserves in dynamic t\&d power markets with
  nondispatchable renewable generation and distributed energy resources,''
  \emph{Proceedings of the IEEE}, vol. 104, no.~4, pp. 807--836, 2016.

\bibitem{6}
``Detailed overview of services technical appendix a : Detailed overview of
  services,'' \emph{RMI}, 2015.

\bibitem{7}
W.~Tushar, J.~A. Zhang, D.~B. Smith, H.~V. Poor, and S.~Thi{\'e}baux,
  ``Prioritizing consumers in smart grid: A game theoretic approach,''
  \emph{IEEE Transactions on Smart Grid}, vol.~5, no.~3, pp. 1429--1438, 2014.

\bibitem{8}
Y.~K. Renani, M.~Ehsan, and M.~Shahidehpour, ``Optimal transactive market
  operations with distribution system operators,'' \emph{IEEE Transactions on
  Smart Grid}, 2017.

\bibitem{9}
Z.~Liu, Q.~Wu, S.~Oren, S.~Huang, R.~Li, and L.~Cheng, ``Distribution
  locational marginal pricing for optimal electric vehicle charging through
  chance constrained mixed-integer programming,'' \emph{IEEE Transactions on
  Smart Grid}, 2016.

\bibitem{10}
S.~Hanif, T.~Massier, H.~B. Gooi, T.~Hamacher, and T.~Reindl, ``Cost optimal
  integration of flexible buildings in congested distribution grids,''
  \emph{IEEE Trans. on Power Systems}, vol.~32, no.~3, pp. 2254--2266, 2017.

\bibitem{11}
S.~Bolognani and F.~D{\"o}rfler, ``Fast power system analysis via implicit
  linearization of the power flow manifold,'' in \emph{Communication, Control,
  and Computing (Allerton), 2015 53rd Annual Allerton Conference on}.\hskip 1em
  plus 0.5em minus 0.4em\relax IEEE, 2015, pp. 402--409.

\bibitem{12}
M.~N. Faqiry and S.~Das, ``Distributed bilevel energy allocation mechanism with
  grid constraints and hidden user information,'' \emph{IEEE Transactions on
  Smart Grid}, 2017.

\bibitem{13}
M.~N. Faqiry, L.~Edmonds, H.~Zhang, A.~Khodaei, and H.~Wu,
  ``Transactive-market-based operation of distributed electrical energy storage
  with grid constraints,'' \emph{Energies}, vol.~10, no.~11, p. 1891, 2017.

\bibitem{14}
M.~{\'A}. Moreno, M.~Bueno, and J.~Usaola, ``Evaluating risk-constrained
  bidding strategies in adjustment spot markets for wind power producers,''
  \emph{International Journal of Electrical Power \& Energy Systems}, vol.~43,
  no.~1, pp. 703--711, 2012.

\bibitem{16}
M.~Lejeune and N.~Noyan, ``Mathematical programming approaches for generating
  p-efficient points,'' \emph{European Journal of Operational Research}, vol.
  207, no.~2, pp. 590--600, 2010.

\bibitem{17}
J.~Li, J.~Wen, S.~Cheng, and H.~Wei, ``Minimum energy storage for power system
  with high wind power penetration using p-efficient point theory,''
  \emph{Science China Info. Sciences}, vol.~57, no.~12, pp. 1--12, 2014.

\bibitem{wu2017stochastic}
H.~Wu, I.~Krad, A.~Florita, B.-M. Hodge, E.~Ibanez, J.~Zhang, and E.~Ela,
  ``Stochastic multi-timescale power system operations with variable wind
  generation,'' \emph{IEEE Transactions on Power Systems}, vol.~32, no.~5, pp.
  3325--3337, 2017.

\bibitem{20}
FERC, ``Electric storage participation in markets operatied by regional
  transmission organization and independent system operators,'' \emph{FERC.
  November}, vol.~17, 2016.

\bibitem{18}
J.~A. Taylor, \emph{Convex optimization of power systems}.\hskip 1em plus 0.5em
  minus 0.4em\relax Cambridge University Press, 2015.

\bibitem{21}
D.~Lew, G.~Brinkman, E.~Ibanez, B.~Hodge, and J.~King, ``The western wind and
  solar integration study phase 2,'' \emph{Contract}, vol. 303, pp. 275--3000,
  2013.

\end{thebibliography}
\end{document}